\documentclass[pra,twocolumn,showpacs,amsmath,amssymb]{revtex4}
\usepackage{color}
\usepackage{graphicx}
\usepackage{amsmath}
\begin{document}

\title{Interacting spin-$1$ bosons in a two-dimensional optical
  lattice}
\author{L. de Forges de Parny$^{1,2}$, F. H\'ebert$^1$,
  V.G. Rousseau$^3$, and G.G. Batrouni$^{1,4,5}$}  
\affiliation{$^1$INLN, Universit\'e de Nice-Sophia Antipolis, CNRS; 
1361 route des Lucioles, 06560 Valbonne, France,}
\affiliation{$^2$Institute of Theoretical Physics, Ecole Polytechnique F\'ed\'erale 
de Lausanne (EPFL), CH-1015 Lausanne, Switzerland,}
\affiliation{$^3$Department of Physics and Astronomy, Louisiana State University,
  Baton Rouge, Louisiana 70803, USA,}
\affiliation{$^4$Institut Universitaire de France, 103 bd Saint-Michel,
75005 Paris, France,}
\affiliation{$^5$Centre for Quantum Technologies, National
University of Singapore; 2 Science Drive 3 Singapore 117542}
\begin{abstract}
  We study, using quantum Monte Carlo (QMC) 
  simulations, the ground state properties of spin-1 bosons trapped in a
  square  optical lattice. The phase diagram is characterized by the mobility of
  the particles (Mott insulating or superfluid phase) and by their magnetic
  properties.  For ferromagnetic on-site interactions, the whole phase
  diagram is ferromagnetic and the Mott insulators-superfluid phase
  transitions are second order.  For antiferromagnetic on-site 
  interactions, spin nematic order is found in the odd Mott lobes
  and in the superfluid phase.  Furthermore, the superfluid-insulator
  phase transition is first or second order depending on whether the
  density in the Mott is even or odd.  Inside the even Mott lobes, we
  observe a singlet-to-nematic transition for certain values of the
  interactions. This transition appears to be first order.
\end{abstract}


\maketitle

\section{Introduction}

Ultracold atoms in optical lattices have been used in the recent
years to explore the physics of well known quantum statistical
discrete models \cite{bloch08}, such as the Bose \cite{Greiner02} and
Fermi \cite{Schneider08,Jordens08} single-band Hubbard models.
Purely optical traps \cite{Stamper1998} allow the trapping of alkali atoms,
such as $^{23}{\rm Na}$, $^{30}{\rm K}$, and $^{87}{\rm Rb}$, in
the $F = 1$ hyperfine state without freezing $F_z$.  The presence of
these spin degrees of freedom allows the study of multi-band condensed
matter Hamiltonians and the interplay between magnetism and
superfluidity \cite{Vengalattore08,Vengalattore10}.

Such systems are governed by extended Hubbard Hamiltonians, with
spin-dependent terms \cite{ho,ohmi98}, able to describe the spinful nature of
the particles.  The on-site spin-spin interaction can be tuned by
using optical Feshbach resonance \cite{Theis} and the nature of the
spin-spin interaction can be either ferromagnetic ($^{87}{\rm Rb}$) or
antiferromagnetic ($^{23}{\rm Na}$) depending on the relative
magnitudes of the scattering lengths in the singlet and quintuplet
channels \cite{Stamperbook}.  In the strong coupling limit and with
integer filling, the spinor Bose-Hubbard model can be mapped onto the
Heisenberg model with biquadratic interactions which exhibits singlet,
dimerized or nematic phases \cite{imambekov04,Kawashima02,Tsuchiya04}.

Several approaches have been used to study the spin-1 bosonic model: mean
field methods \cite{pai08,tsuchiya05}, variational Monte Carlo
\cite{toga}, analytical \cite{imambekov04,Katsura}, strong coupling expansion \cite{Kimura13}, Density Matrix Renormalization Group (DMRG)
\cite{Rizzi05_dmrg,Bergkvist06_dmrg}, and quantum Monte Carlo simulations in 1D
\cite{apaja06,batrouni2009}.  The overall picture emerging at zero
temperature is that, as expected, the system adopts Mott-insulating
(MI) phases, when the filling is commensurate with the lattice size
and for large enough repulsion between particles, and a superfluid
phase (SF) otherwise.  The richness of these systems comes from the
magnetic behavior of these phases that changes dramatically with the
sign of the spin-dependent interaction parameter $U_2$.

For negative spin-dependent interaction $U_2<0$ (see below) corresponding to the
ferromagnetic case, a 1D QMC study \cite{batrouni2009} shows that all
Mott lobes shrink, and eventually disappear, as one increases $|U_2|$.
The whole phase diagram is ferromagnetic and the same is expected in
larger dimension. This was confirmed by a recent theorem that shows
that the ground state of such a system is always ferromagnetic
\cite{Katsura}.

The positive $U_2$ antiferromagnetic case exhibits a richer phase
diagram.  For even densities, $\rho$, the Mott lobes are expected to
have spin singlets on every site but a first order transition to a
nematic state is predicted for $U_2/U_0 < 0.025$ in two
dimensions \cite{imambekov04,demler02,snoek04}.  For odd $\rho$, a nematic order
is also expected in two dimensions, depending on the interaction $U_2$
\cite{imambekov04} (in one dimension, this nematic phase is changed
into a dimerized phase \cite{Rizzi05_dmrg,batrouni2009}).  The superfluid phase is
expected to be polarized: superfluidity is carried either by the $F_z
= 0$ component or by the $F_z=\pm1$ components \cite{pai08}.  The
theorem previously mentioned \cite{Katsura} proves that the global magnetization is
zero in the whole phase diagram in the subspace of zero total spin along the $z$ axis ($F_{{\rm tot},z}=0$). 
This is compatible
with the nematic or dimerized phases observed in other studies.

As for the nature of the MI-SF phase transitions, it was predicted to be
affected by the spin-spin interactions \cite{pai08,tsuchiya05}. For $\rho=2$ it is
argued to be first-order for small positive $U_2$ (and second order in
all other cases) whereas the transition for $\rho=1$ should always be
second order.  The one-dimensional QMC investigation
\cite{batrouni2009} presented evidence of singlet and dimerized phases
but found only second order transitions, which is expected in one
dimension. Recent 1D and 2D QMC studies \cite{deforges10,deforges11} of a similar
system with only two species showed a first order transition for even
lobes occurring in two dimensions, as predicted by previous mean-field studies \cite{kruti04,kruti05}, but not in the one-dimensional case.

The purpose of this paper is to extend the QMC study of the spin-1
system to the two-dimensional square lattice at zero temperature.  The
paper is organized as follows: in Sec. II, we will introduce the model
and QMC techniques used to study it.  Secs. III and IV will be devoted
to the presentation of the results obtained for the $U_2< 0$ and
$U_2>0$ cases, respectively.  In Sec. V, we will summarize these
results and give some final remarks.

\section{Spin-1 model\label{section2}}

We consider a system of bosonic atoms in the
hyperfine state $F=1$ characterized by the magnetic quantum number
$F_z=0,\pm1$. When these atoms are loaded in an optical lattice,
the system is governed by the Bose-Hubbard Hamiltonian \cite{ho, demler02}:
\begin{eqnarray}
  \nonumber
  \mathcal H&=&-t\sum_{\sigma,\langle \bf r,\bf r' \rangle} \left (a^\dagger_{\sigma \bf r}
    a^{\phantom\dagger}_{\sigma \bf r'} + {\rm h.c.}\right ) + \frac{U_0}{2}\sum_{\bf r}
  {\hat n}_{ \bf r} \left ( {\hat n}_{\bf r}-1\right )\\
  &&+ \frac{U_2}{2}\sum_{\bf r} \left({\bf F}_{\bf r}^2-2 {\hat n}_{\bf r} \right ),
\label{ham1}
\end{eqnarray}
where operator $a^{\phantom\dagger}_{\sigma {\bf r}}$ ($a^ \dagger_{\sigma {\bf r}}$)
annihilates (creates) a boson of spin $\sigma =-1,0,1$ (also denoted
$-,0,+$ when needed) on site ${\bf r}$ of a periodic square lattice of
size $L\times L$.

The first term in the Hamiltonian is the kinetic term which allows
particles to hop between neighboring sites $\langle {\bf r,r'} \rangle$.
The hopping parameter $t = 1$ sets the energy scale and the number operator
$\hat{n}_{ {\bf r}} \equiv \sum_{\sigma} {\hat n}_{ \sigma \bf r} =
\sum_{\sigma} a^\dagger_{\sigma {\bf r}} a^{\phantom{\dagger}}_{\sigma
  {\bf r}}$ counts the total number of bosons on site $\bf r$.
$N_\sigma = \sum_{\bf r} n_{\sigma {\bf r}}$ will denote the total
number of $\sigma$ bosons, $\rho_\sigma = N_\sigma/L^2$ the
corresponding density, and $\rho$ the total density.  The operator
${\bf F}_{\bf r} = (F_{x, {\bf r}}, F_{y,{\bf r}}, F_{z,{\bf r}})$ is
the spin operator where $ F_{\alpha,{\bf r}}=\sum_{\sigma,\sigma' }
a^\dagger_{\sigma \bf r} J_{\alpha,\sigma \sigma'} a_{\sigma' \bf r}$,
$\alpha = x,y,z$ and the $J_{\alpha, \sigma \sigma'}$ are standard
spin-1 matrices.  Once developed, the ${\bf F}^2_{\bf r}$ term exhibits
contact interaction terms and conversion terms between the species.  The
parameters $U_0$ and $U_2$ are the on-site spin-independent and
spin-dependent interaction terms.  The nature of the bosons will give
the sign of $U_2$ and consequently the nature of the on-site spin-spin
interaction whereas the spin independent part is always repulsive,
$U_0 > 0$.  The $U_2<0$ case (\rm{e.g.} $^{87}{\rm Rb}$) will maximize
the local magnetic moment $F^2(0)\equiv \sum_{\bf r}\langle
{\bf F}_{\bf r}^2 \rangle/ L^2$.  Therefore, this is referred to as the
``ferromagnetic" case whereas the $U_2>0$ case (\rm{e.g.} $^{23}{\rm
  Na}$) which favors minimum on-site spins is referred to as
``antiferromagnetic".

To simulate this system, we used the Stochastic Green Function
algorithm (SGF) \cite{SGF} with directed updates \cite{directedSGF},
an exact Quantum Monte Carlo technique that allows canonical or grand
canonical simulations of the system as well as measurements of
many-particle Green functions.  In particular, this algorithm can
simulate efficiently the spin-flip term $ a^\dagger_{0 \bf
  r}a^\dagger_{0 \bf r}a^{\phantom\dagger}_{- \bf r} a^{\phantom\dagger}_{+ \bf r} + {\rm h.c.} $ which
appears in the development of the ${\bf F}_{\bf r}^2$ term.  In this
work we used mostly the canonical formulation where the total density
$\rho$ is conserved whereas the individual densities $\rho_\sigma$
fluctuate. The QMC algorithm also conserves the total spin along
$z$, $F_{{\rm tot},z}= N_{+} - N_{-}$, which adds a constraint to the
canonical one. The value of $F_{{\rm tot},z}$ in a given canonical simulation
is then fixed by the choice of the initial numbers of particles.
Due to this constraint, the magnetic physical
quantities, involving $F_{\alpha,{\bf r}}$ operators, are identical
for the $x$ and $y$ axes but may be different for the $z$ axis. The
initial symmetry, where all three axes should behave identically, is
broken in our simulations.
Using this algorithm we were able to simulate the system
reliably for clusters going up to $L\times L=12\times 12$ and 
$\beta= 2 L/t$. We generally compared the 
results for sizes $L=8, 10$, and 12 to check that there was no appreciable
difference in the behaviour of the system
within the limit of the precision we could achieve.

At zero temperature, we relate the density to the chemical potential
$\mu$ using
\begin{equation}
\mu(N)=E(N+1)-E(N), \label{mu_deltaE}
\end{equation}
where $N$ is the total number of bosons, and $E= \langle \mathcal H
\rangle $ is the average energy, equal to the free energy in the
ground state.

The superfluid density
is given by \cite{roy}
\begin{equation}
\rho_s= \frac{\langle W^2 \rangle}{4t\beta},
\label{spinone_rhosc2D}
\end{equation}
where the total winding number $W=W_{-}+W_0+W_{+}$ is a topological
quantity, and $\beta=1/k_BT$.  The analysis of the magnetic structure
requires the equal-time spin-spin correlation functions
\begin{equation}
  F_{\alpha \alpha}( {\bf R}) = \frac{1}{L^2} \sum_{\bf r} \langle
  F_{ \alpha, {\bf r}} F_{ \alpha, {\bf r+R}}  \rangle.  \label{Fxx}
\end{equation}
The global magnetization is given by
\begin{equation}
 F^2_{\rm tot} =  L^2\sum_{\alpha, {\bf R}}   F_{\alpha \alpha}({\bf
   R}) 
\end{equation}
and the magnetic structure factor by 
\begin{equation}
S({\bf k}) = \frac{1}{L^2}\sum_{\alpha,\bf R} e^{i{\bf k\cdot R}}
F_{\alpha \alpha}({\bf R}) 
\end{equation}
where ${\bf k} = (k_x,k_y)$ and $k_{x,y}$ are integer multiples of $2 \pi /
L$.  Other important quantities are the diagonal components of the
nematic order parameter traceless tensor $\mathbf{Q}$ given by
\cite{demler02} 
\begin{equation}
 Q_{\alpha \alpha}= \langle   F_{\alpha \alpha}(0)-\frac{1}{3}F^2(0)
 \rangle.    \label{Qxx} 
\end{equation}
These are the indicators of the on-site spin isotropy if $Q_{\alpha
  \alpha}=0$ for all $\alpha$,  and of the spin anisotropy
characteristic of the nematic order if $Q_{\alpha \alpha} \ne 0$ and if,
at the same time, no magnetic order is present. As $ Q_{xx}= Q_{yy}$ in our
simulations, then $2Q_{xx}+Q_{zz}=0$ and it is sufficient to calculate
$Q_{zz}$. Since nematic order can develop along any axis and as
there is not always an ergodicity breaking in MC simulations on finite
clusters, $Q_{zz}$ can change sign and average to zero even if it is
non-zero in each individual measurements. To avoid such an effect, we use
the absolute value of the sums of spin operators in $Q'_{zz}= \langle
|F_{zz}(0)-\frac{1}{3}F^2(0)| \rangle$.

We also calculate the Green functions
\begin{equation}
  G_\sigma({\bf R}) =\frac{1}{2L^2}\sum_{\bf r} \langle a^\dagger_{\sigma {\bf
      r+R}}a^{\phantom{\dagger}}_{\sigma {\bf r}} + a^\dagger_{\sigma {\bf
      r}}a^{\phantom{\dagger}}_{\sigma {\bf
      r+R}}\rangle,
\label{green1}
\end{equation}
which measure the phase coherence of particles.  The anticorrelated
motions of particles, which govern the dynamics inside Mott lobes, as
the particles of different types exchange their positions
are described by the two-particle anticorrelated Green functions
\begin{equation}
  G_{a}(\sigma\sigma',{\bf R})  = \frac{1}{2L^2}   \sum_{{\bf r}} \langle    
  \hat{a}_{\sigma {\bf r+R}}^{\dagger}    
  \hat{a}_{\sigma' {\bf r+R}}^{\phantom\dagger}    
  \hat{a}_{\sigma' {\bf r}}^{\dagger}  
  \hat{a}_{\sigma {\bf r}}^{\phantom\dagger} + {\rm h.c.}  \rangle. \label{Gapm} 
\end{equation}
\\
If perfect phase coherence is established by means of particle
exchange, $G_{a}(\sigma \sigma',{\bf R}) $ reaches its limiting upper
value of $\rho_\sigma \rho_{\sigma'} $ at long distances ${\bf R}$.
Due to its definition, $G_{a}(\sigma \sigma',{\bf R}) =G_\sigma({\bf
  R}) G_{\sigma'}({\bf R})$ if there is no correlation between the
movements of particles of spin $\sigma $ and $\sigma'$.

\section{Ferromagnetic case: ${U_2<0}$}

We start with the ferromagnetic case, ${U_2<0}$.  Several approaches
have been used to study this case.  A recent theorem by Katsura and
Tasaki, independent of the lattice dimension, shows that the ground
state universally exhibits saturated ferromagnetism both in superfluid
and Mott insulator phases \cite{Katsura}.  Mean field analyses
predicted the same magnetic behavior and continuous MI-SF phase transitions
\cite{pai08,tsuchiya05}.  The 1D investigation \cite{batrouni2009}, by means of
QMC simulations, showed that the magnetic local moment on a site with
$n$ bosons is maximized $F^2(0)=n(n+1)$ and that the whole phase
diagram is ferromagnetic, in agreement with the above theorem.  In the
2D system, studied with variational Monte Carlo at
$\rho=1$ \cite{toga}, Mott and superfluid phases were also found to be
ferromagnetic.  In the first subsection, we calculate the 2D phase
diagram in the $(\mu/U_0,t/U_0)$ plane at a fixed ratio $U_2/U_0=-0.1$
and we discuss the phase transitions in the second subsection.

\subsection{Phase diagram} 

\begin{figure}[h]
	\includegraphics[height=6.8  cm]{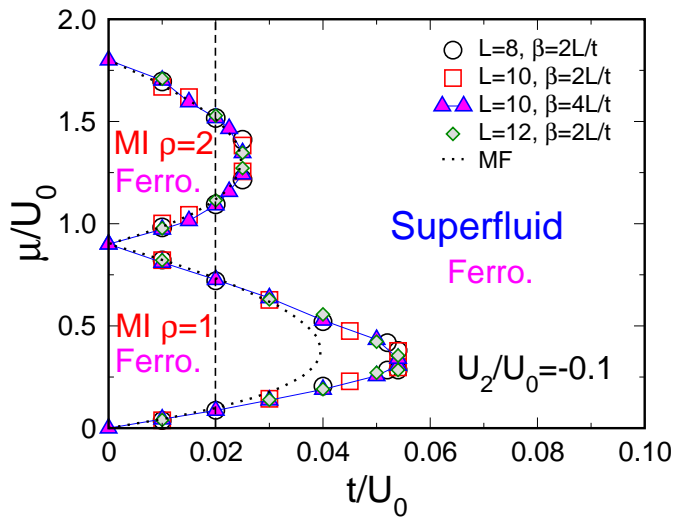}
        \caption {(Color online) $F_{{\rm tot},z}=0$ phase diagram at $\beta=2L/t$ or
          $\beta=4L/t$ for $U_2/U_0=-0.1$ and several linear system
          sizes $L$. All phases are ferromagnetic and all transitions
          are continuous.  The dashed vertical line
          indicates where the cut in
          Fig.~\ref{spinone_rhovsmu_MCQ_F_coupediagphase_Utzm0p1_tsurUz_0p020_L8beta32}
          was taken. The dotted line corresponds to the MF phase diagram obtained with the technique
used in \cite{pai08}.  }
\label{Spinone_QMC_diag_phase_ferro_Uz1_Utm0p1Uz}
\end{figure}

In the no-hopping limit, $t/U_0 \to 0$, the system is in a MI phase
with $n$ bosons per site with energy per site $\varepsilon(n) =
n(n-1)(U_0+U_2)/2$ when $\mu$ satisfies the condition $
(n-1)(U_0+U_2) < \mu < n(U_0+U_2)$.  Therefore, the bases of
both the odd and even Mott lobes, given by $\Delta \mu
/U_0=1-|U_2|/U_0$, shrink with increasing $|U_2|$ and disappear
completely when $|U_2|=U_0$.  The on-site energy is minimized by
maximizing the quantum number $F$ corresponding to $F^2(0) = F(F+1)$,
that is by adopting any of the $F=1$ triplet states for $\rho=1$ and
any of the $F=2$ quintuplet states for $\rho=2$.  Thus, in the $t/U_0
\to 0$ limit, the Mott phases are highly degenerate.  As $t/U_0$
increases, the MI regions are reduced and eventually disappear, giving
the familiar Mott lobes
(Fig.~\ref{Spinone_QMC_diag_phase_ferro_Uz1_Utm0p1Uz}).  Outside the
MI region the system is superfluid.
Note that the shape of the lobes is very different from the 1D case
\cite{batrouni2009} where the lobes end in cusps.
The MF results obtained with the technique used in \cite{pai08}
are in very good agreement with QMC results at large interactions, as expected.
The tip of the $\rho=2$ is also well reproduced by the MF approximation whereas
there is a quantitative difference between QMC and MF 
for $\rho=1$.

\begin{figure}[h]
	\includegraphics[height=6.8  cm]{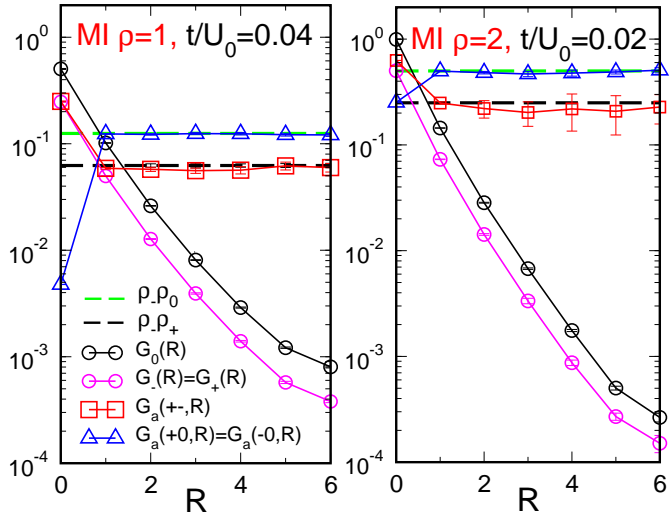}
        \caption {(Color online) The single-particle and the
          anticorrelated Green functions for $U_2/U_0=-0.1$, $L=12$
          and $\beta=2L/t$ as a function of distance $R$ in the Mott insulator
          phases with $\rho=1$ and $\rho=2$.  Individual movements of
          particles are strongly suppressed, as shown by the
          exponential decay of $G_0$ and $G_-=G_+$, whereas perfect
          anticorrelated movements of particles remain, which is shown
          by the plateau in $G_a(+-) \to \rho_-\rho_+$ and
          $G_a(+0)=G_a(-0) \to \rho_\pm\rho_0$ when $R \to L/2$.  }
\label{Spinone_ferro_Green_MI_rho1_2}
\end{figure}

In the Mott insulator phases, the superfluid density is zero and the
densities are $\rho_0=2 \rho_+=2\rho_-=\rho/2$.  The degenerate states
of the MI phases in the $t/U_0 \to 0$ limit are coupled by
second-order contributions from the hopping term which exchange the
position of different particles. The degeneracy is, then, lifted by
establishing a phase coherence of such exchanges \cite{Kuklov2003,
  svistunov2, deforges11}.  Figure \ref{Spinone_ferro_Green_MI_rho1_2}
shows single-particle and anticorrelated Green functions in the MI
phases for $\rho=1$ and $\rho=2$.  As expected in the MI phase, we
see that the individual Green functions $G_\sigma({\bf R})$ decay
exponentially to zero with distance ${\bf R}$.  Figure
\ref{Spinone_ferro_Green_MI_rho1_2} shows that the
$G_a(\sigma\sigma',{\bf R})$ quickly saturate to a constant value at
large separations, indicating coherent exchange moves in the MI phase
for both $\rho=1$ and $\rho=2$.  Furthermore, the plateau reached by
this function takes on its maximum possible value at large distances,
$\rho_\sigma\rho_{\sigma'}$, showing perfect phase coherence.  The
equalities $G_-({\bf R})=G_+({\bf R})$ and $G_{a}(+0,{\bf
  R})=G_{a}(-0,{\bf R})$ are due to the $F_{{\rm tot},z}=0$ constraint.
As $\rho_0 > \rho_\pm$, the Green functions involving spin-0 particles
($G_0({\bf R})$ and $G_a(\pm 0,{\bf R})$) are larger than Green functions
involving only $\pm 1$ particles, which will also be true in
the superfluid case.
 
\begin{figure}[h]
	\includegraphics[height=6.8  cm]{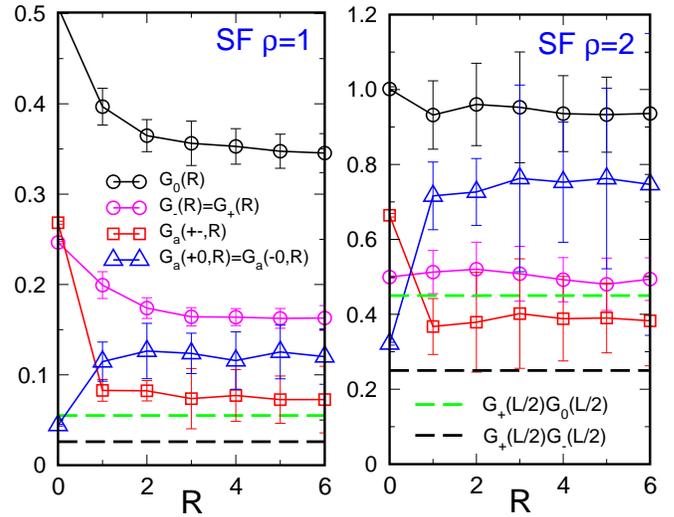}
        \caption {(Color online) The single-particle and the
          anticorrelated Green functions for $U_2/U_0=-0.1$, $L=12$,
          $\beta=2L/t$ with $t/U_0=0.10$ as a function of distance  $R$ in the
          superfluid phase with $\rho=1$ and $\rho=2$.  The long-range
          order of the Green functions indicates both individual and
          anticorrelated movements of particles.  }
\label{Spinone_ferro_Green_SF_rho1_2}
\end{figure}

In the superfluid phase, we observe the same density distribution
$\rho_0=2 \rho_+=2\rho_-=\rho/2$.  Figure
\ref{Spinone_ferro_Green_SF_rho1_2} shows phase coherence both for
exchange moves and for individual movements of particles.  The
saturation of the single-particle Green functions $G_\sigma({\bf R})$
to a constant value at large separations is the signature of the
existence of a Bose-Einstein condensate as the condensed fraction is
given by $\rho({\bf k}=0)=\sum_{\sigma}\rho_\sigma({\bf k}=0)$, where
the $\rho_\sigma({\bf k})$ are Fourier transforms of single-particle
Green functions.  For both densities, $\rho=1$ and $\rho=2$, the three
species remain correlated, of course, since the system is still in a
strongly interacting regime $(t/U_0=0.10)$. This correlation can be
inferred from the fact that $G_{a}(+-,{\bf R}) > G_+({\bf R}) G_-({\bf
  R})$ and $G_{a}(\pm 0,{\bf R}) > G_\pm({\bf R}) G_0({\bf R})$ at
large separations.  This means that, while particles can move
independently, exchanges of different particles are still present.
This is typical of a strongly correlated superfluid where different
kinds of phase coherence can be observed: phase coherence of the
individual particles, but also, at the same time, phase coherence of
exchange moves of particles.\\

\begin{figure}[!h]
\begin{center}
	\hspace{0.8 cm}\includegraphics[height=6.5 cm]{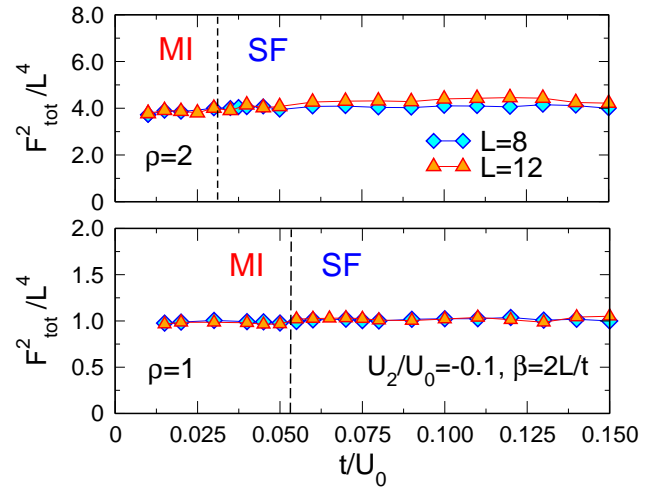}
        \caption {(Color online) The normalized total magnetization
          $F^2_{\rm tot}$ versus $t/U_0$ at fixed total
          density $\rho=1,2$. The dashed vertical lines indicate the
          critical values $t_c/U_0$ where the MI-SF phase transitions
          occur.}
\label{BISspinone_F_QMC_Nematique_transition_rho12_Utz0p1}
\end{center}
\end{figure}

\begin{figure}[!h]
\begin{center}
  \hspace{-0.1 cm}\includegraphics[height=6.5 cm]{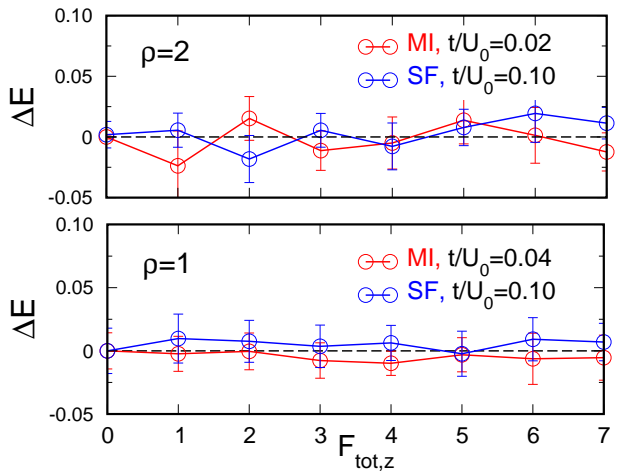}
  \caption{(Color online) The energy difference $\Delta E = E(F_{{\rm
        tot},z})-E(F_{{\rm tot},z}=0)$ versus $F_{{\rm tot},z}$ in the
    MI and SF phases for $\rho=1,2$ with $U_2/U_0=-0.1$, $L=4$ and
    $\beta=6L/t$.  $\Delta E$ is zero, in the error bars limit,
    indicating that the ground states are $(2N + 1)$-fold degenerate.
  }
\label{spinone_E_L4b120Utz0pm1_MI_SF_rho1_2}
\end{center}
\end{figure}

We now focus on the magnetic properties of MI and SF phases for
$U_2/U_0=-0.1$.  The ground state exhibits saturated ferromagnetism in
the whole phase diagram, as predicted by Katsura and Tasaki's theorem
\cite{Katsura} and previously observed in \cite{batrouni2009,toga}.
In both MI and SF phases with integer density $\rho=1,2$ we measure
the largest possible spin on-site, given by $F^2(0)=\rho(\rho+1)$.
Furthermore, the ground state has the maximum possible total spin
$F^2_{\rm tot}=N(N+1)$, thus $F^2_{\rm tot}/L^4 = \rho(\rho + 1/L^2)
\simeq \rho^2$
(Fig.~\ref{BISspinone_F_QMC_Nematique_transition_rho12_Utz0p1}) in the
whole phase diagram, in the superfluid phase as well as in the Mott
phase.  Figure~\ref{spinone_E_L4b120Utz0pm1_MI_SF_rho1_2} shows that
the ground state energy $E(F_{{\rm tot},z})$ is independent of the
$F_{{\rm tot},z}$ chosen in the simulation, in the MI or in the SF
phase with $\rho=1$ or 2. As mentionned earlier, $F_{{\rm tot},z}$
is imposed by the initial values of $N_+$ and $N_-$ used in a given simulation.
The ground states are then $(2N + 1)$-fold
degenerate, as expected if $F^2_{\rm tot} = N(N+1)$.  This also shows
that the constant $F_{{\rm tot},z}$ we impose just select one of these
$2N+1$ states and does not alter the physics.

\begin{figure}[!h]
\begin{center}
  \hspace{0 cm}\includegraphics[height=6. cm]{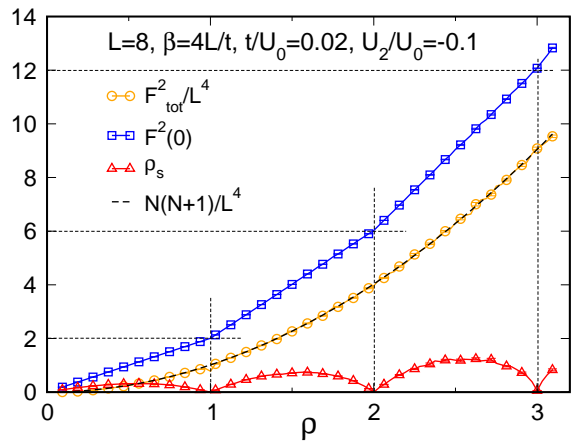}
\caption {(Color online) Normalized total magnetization $F^2_{\rm
    tot}/L^4$, local magnetic moment $F^2(0)$, and superfluid density
  $\rho_s$ versus the density $\rho$ associated with the vertical slice
  at $t/U_0=0.02$ of the phase diagram
  Fig.~\ref{Spinone_QMC_diag_phase_ferro_Uz1_Utm0p1Uz}. The total
  magnetization $F^2_{\rm tot}$  takes the maximum possible value $N(N+1)$.
\label{spinone_rhovsmu_MCQ_F_coupediagphase_Utz0p1_tsurUz_0p040_L8beta16_2}
}
\end{center}
\end{figure}

Figure~\ref{spinone_rhovsmu_MCQ_F_coupediagphase_Utz0p1_tsurUz_0p040_L8beta16_2}
shows $F^2_{\rm tot}/L^4$, $F^2(0)$, and the superfluid density
$\rho_s$ versus the density $\rho$. It shows that $F^2_{\rm tot} =
N(N+1)$ and that the magnetic properties are not influenced by the
superfluid or solid nature of the underlying phases.  The linear step
by step evolution of $F^2(0)$ is simply explained by the strong
on-site repulsion $U_0$. For a density $\rho$ we define $\rho_i$, the
integer part of the density (for $2 \le \rho < 3$, $\rho_i = 2$). To
minimize the repulsion cost, a system with density $\rho$ will have
the minimal number of sites occupied by $\rho_i+1$ particles, that is
$(\rho-\rho_i)L^2$. These sites contribute $(\rho_i + 1)(\rho_i+2)$ to
the local moment and the remaining $(\rho_i+1-\rho)L^2$ sites occupied
by $\rho_i$ particles contribute $\rho_i(\rho_i+1)$.  Averaging these
contributions yields the observed dependence which is $F^2(0) =
(\rho_i+1)(2\rho - \rho_i)$.

\subsection{Phase transitions} 
 
\begin{figure}[h]
  \includegraphics[height=6.5 cm]{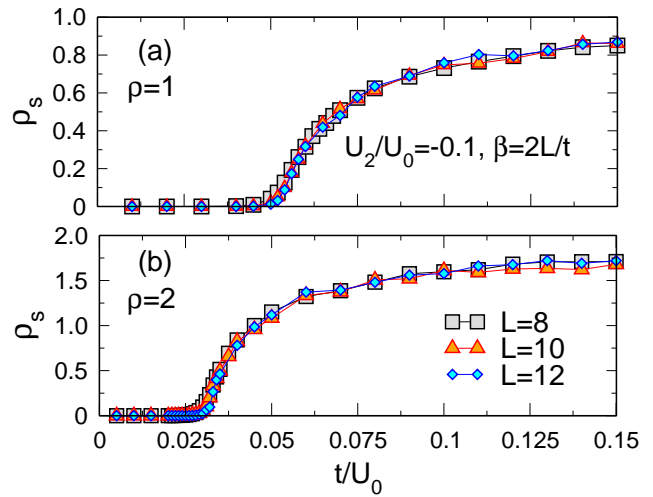}
  \caption {(Color online) Superfluid density $\rho_s$ as a function
    of $t/U_0$ in the first (a) and second (b) Mott lobes. No evidence
    of a discontinuous jump can be seen as the system size increases;
    transitions are second-order.}
\label{spinone_F_QMC_transition_rho1et2_Utz0p1}
\end{figure}

Figure~\ref{spinone_F_QMC_transition_rho1et2_Utz0p1} shows the
evolution of the superfluid density $\rho_s$ at the MI-SF transition
at fixed total density while varying $t/U_0$, in other words the
transition at the tip of the Mott lobe.  All quantum phase transitions
appear to be continuous and there are no signs of possible
discontinuities in the superfluid density at the transition between
the MI and SF phases.  The finite size effects are small and the
transitions appear continuous, which means that they should be in the universality class of the
three-dimensional XY model, which is expected from general scaling
arguments \cite{fisher}.  These results are in agreement with the mean
field analysis \cite{pai08}.

\begin{figure}[!h]
\begin{center}
  \hspace{0. cm}\includegraphics[height=6.5 cm]{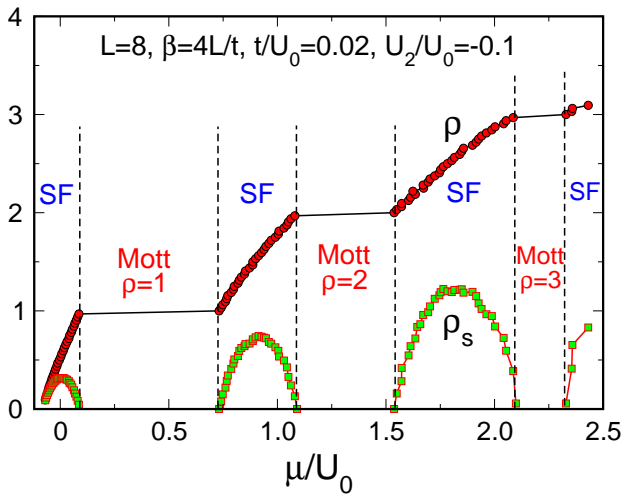}
  \caption{(Color online) The total density, $\rho$ as a function of
    the chemical potential, $\mu$, showing the first three
    incompressible Mott phases. Also shown is the superfluid density
    $\rho_{\rm s}$ in the compressible phases. All the transitions are
    second-order.  }
\label{spinone_rhovsmu_MCQ_F_coupediagphase_Utzm0p1_tsurUz_0p020_L8beta32}
\end{center}
\end{figure}

Figure~\ref{spinone_rhovsmu_MCQ_F_coupediagphase_Utzm0p1_tsurUz_0p020_L8beta32}
shows a cut in the phase diagram as a function of $\mu$ at $t/U_0 =
0.02$ (dashed vertical line in
Fig.~\ref{Spinone_QMC_diag_phase_ferro_Uz1_Utm0p1Uz}) and clearly
exhibits the first three incompressible Mott plateaux where the
superfluid density vanishes.
Here again, we do not observe discontinuities in $\rho$ or $\rho_s$
and conclude that the transitions are continuous.\\ 

\section{Antiferromagnetic case: $U_2>0$}

\subsection{Phase diagram}

In the no-hopping limit, $t/U_0 \to 0$, the system is in a MI phase
with $n$ bosons per site when $\mu$ satisfies the condition
$(n-1)U_0<\mu< n U_0-2U_2$ for odd density and $U_0(n-1)- 2U_2<\mu< n
U_0$ for even density.  Therefore, the base of the Mott lobes for odd
filling is $\Delta \mu/U_0=1-2U_2/U_0$, whereas it is $\Delta
\mu/U_0=1+2U_2/U_0$ for even filling.  The even lobes grow at the
expense of the odd ones, which disappear entirely for $U_2/U_0=0.5$.

\begin{figure}[!h]
 \includegraphics[height= 6.5 cm]{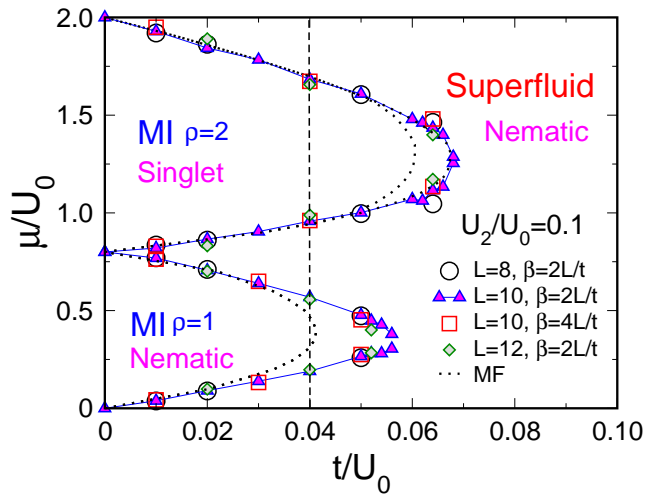}
 \caption {(Color online) $F_{{\rm tot},z}=0$ phase diagram at
   $\beta=2L/t$ or $\beta=4L/t$ for $U_2/U_0=0.1$ and several linear
   system sizes $L$.  In the whole phase diagram, the global
   magnetization is zero, $F^2_{\rm tot}=0$.  In the $\rho=2$ MI, the
   on-site spin is zero, $F^2(0)=0$, the pair of particles forming a
   singlet.  The $\rho=1$ MI and the superfluid phase exhibit
   nematic order and the system is essentially composed of spin $\pm1$
   particles.  The MI-SF phase transition with $\rho=1$ ($\rho=2$) is
   second (first)-order.  The dashed vertical line indicates where the
   cut in
   Fig.~\ref{spinone_rhovsmu_MCQ_AF_coupediagphase_Utz0p1_tsurUz_0p040_L8beta16}
   was taken. The dotted line corresponds to the MF phase diagram obtained with a technique
similar to \cite{pai08}, the two lines at the tip of the $\rho=2$ lobe correspond
a SF-MI coexistence region.}
 \label{Spinone_QMC_diagphase_antiferro_Ut0p1Uz}
 \end{figure}

The on-site energy is minimized by minimizing $F^2(0)$, that is by
adopting one of the $F=1$ triplet states for $\rho=1$ and a unique
$F=0$ singlet state for $\rho=2$, which is given at site ${\bf r}$ by 
\begin{eqnarray}
\nonumber
|\Phi_{{\rm sg},{\bf r}}\rangle &=&A^\dagger_{{\rm sg},{\bf
r}}|0\rangle\\
&=&\frac{1}{\sqrt{6}}\left(2 a^\dagger_{+{\bf r}} a^\dagger_{-{\bf r}} -
a^\dagger_{0{\bf r}} a^\dagger_{0{\bf r}}\right)|0\rangle
\label{sing}
\end{eqnarray}
This defines the singlet creation operator. Thus, in the $t/U_0 \to 0$
limit, the MI $\rho=1$ phase is highly
degenerate (as in the $U_2 < 0$ case) whereas the MI $\rho=2$ state is
unique.  As $t/U_0$ increases, quantum fluctuations cause the MI region
to shrink and eventually disappear. The system is superfluid outside
the MI region
(Fig.~\ref{Spinone_QMC_diagphase_antiferro_Ut0p1Uz}). As in the $U_2 < 0$ 
case, the MF approximation \cite{pai08} reproduces well the base of the Mott lobes
and the tip of the $\rho=2$ lobe, whereas there is a significant difference
for the position of the tip of the $\rho=1$ lobe.
 Throughout phase diagram the global magnetization is zero,
 $F^2_{\rm tot}=0$, in the subspace $F_{{\rm tot},z}=0$.  Despite
 this, we will see below that the two Mott lobes exhibit distinct
 behavior: the lifting of degeneracy in the first lobe will yield a
 nematic phase whereas the second lobe exhibits in most cases a
 singlet phase, with a transition to a nematic phase under certain
conditions~\cite{imambekov04}.

\begin{figure}[h]
  \includegraphics[height=6.8 cm]{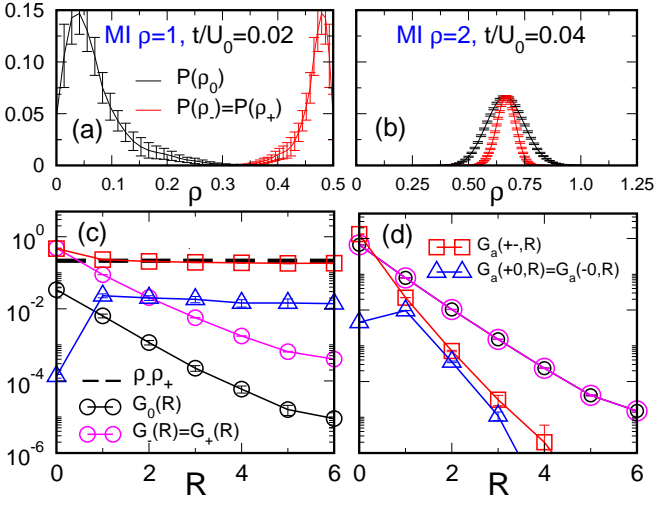}
  \caption {(Color online) (a) and (b): Density distributions of
    spin-$0$ and spin-$\pm1$ particles in the $\rho=1,2$ MI phases and
    $U_2/U_0=0.1$, $L=12$, $\beta=2L/t$.  In the first lobe, there are
    almost only $F_z = \pm 1$ particles, whereas the densities of the
    different species are equilibrated for $\rho=2$.  (c) and (d): The
    single-particle and the anticorrelated Green functions in the same
    cases.  In both lobes, individual movements of particles are
    strongly suppressed.  The first MI lobe (c) exhibits perfect
    anticorrelated movements of particles: $G_a(+-, R \to L/2) \to
    \rho_-\rho_+$, which is not present in the second (d).  }
\label{Spinone_antiferro_Green_MI_rho1_2}
\end{figure}

 For the $\rho=1$ MI, the density distribution
 (Fig.~\ref{Spinone_antiferro_Green_MI_rho1_2} (a)) shows that the
 system is essentially composed of the $F_z=\pm1$ components. 
  These results agree with Ref.~\cite{toga}.  
This is in contradistinction to the $U_2<0$ case
where $\rho_0$ was larger than $\rho_\pm$. Accordingly, in the present
case, the largest Green functions will be those involving $\pm 1$ particles.
As observed for the $U_2<0$
 case, Fig.~\ref{Spinone_antiferro_Green_MI_rho1_2}~(c) shows that
 $G_{a}(+-,{\bf R})$ quickly saturates to its maximum possible value
 $\rho_-\rho_+$ at large distances and is larger than $G_{a}(\pm0,{\bf R})$. This indicates the presence of
 coherent exchange moves, whereas the individual $G_\sigma({\bf R})$
 decay exponentially with distance.

In the $\rho=2$ MI phase, the system is frozen in a state with two
particles per site, which form a singlet pair.  In this phase, the
singlet density, defined as \begin{eqnarray}
\label{spinone_singletdensity}
\rho_{\rm sg}=
\frac{1}{L^2}\sum_r\langle  {\hat A}_{{\rm sg},{\bf r}}^{\dagger}
{\hat A}_{{\rm sg},{\bf r}}^{\phantom\dagger}\rangle,
\end{eqnarray}
is approximately equal to 1 (see, for example,
Fig.~\ref{spinone_AF_FF_rhos_QMC_transition_rho2_Utz0p1}). The
densities are, accordingly, $\rho_+=\rho_-=\rho_0=2/3$ as shown in
Fig.~\ref{Spinone_antiferro_Green_MI_rho1_2} (b) where the density
distributions are all centered around $2/3$.  All the Green functions
decay exponentially to zero with distance
(Fig.~\ref{Spinone_antiferro_Green_MI_rho1_2} (d)):
the anticorrelated movements are absent in this phase because there is
no degeneracy to lift when $t/U_0\neq0$.\\

\begin{figure}[h]
  \includegraphics[height=6.8 cm]{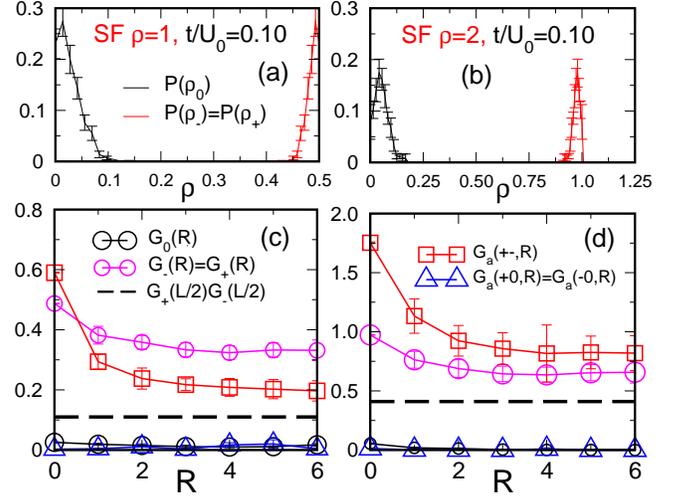}
  \caption {(Color online) (a) and (b): Density distributions of
    spin-$0$ and spin-$\pm1$ particles in the SF phase for $\rho=1,2$
    and $U_2/U_0=0.1$, $L=12$, $\beta=2L/t$.  For both $\rho=1,2$, the
    peak of spin-$\pm1$ distribution approaches the density $\rho/2$,
    while the spin-0 one approaches $0$.  (c) and (d): The
    single-particle and the anticorrelated Green functions in the SF
    phase in the same cases.  For both $\rho=1,2$, individual
    movements and anticorrelated movements of spin-$\pm1$ particles
    are present.  }
\label{Spinone_antiferro_Green_SF_rho1_2}
\end{figure}

In the superfluid phase, Fig.~\ref{Spinone_antiferro_Green_SF_rho1_2}
(a) and (b) show that the peak of spin-$\pm1$ distribution approaches
the density $\rho/2$, while the peak for spin-0 approaches $0$ for
both densities $\rho=1,2$.  Consequently, the superfluid is mostly
composed of $F_z = \pm 1$ particles, same as in the $\rho=1$ Mott
phase.  This density distribution, observed in Ref.~\cite{toga} for
$\rho=1$, differs from the mean field one which predicts that
superfluidity is carried either by the $F_z = 0$ component or by the
$F_z=\pm1$ components \cite{pai08}.  Furthermore, Figure
\ref{Spinone_antiferro_Green_SF_rho1_2} (c) and (d) show a phase
coherence both for exchange moves and for individual movements of
spin-$\pm1$ particles in a way similar to the $U_2 < 0$ case.
 
\begin{figure}[!h]
\begin{center}
  \hspace{-0.1 cm}\includegraphics[height=6.1 cm]{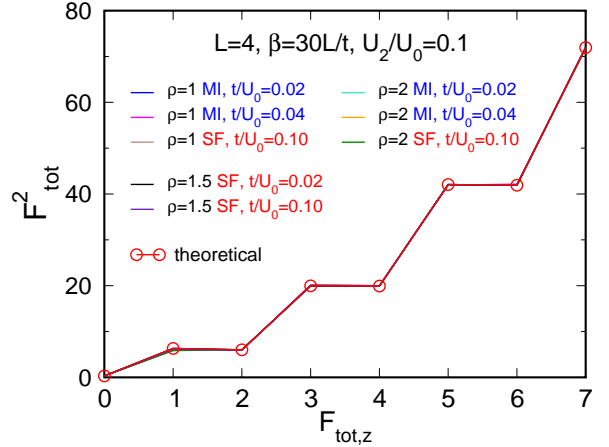}
                \caption{(Color online) Global magnetization versus
                  $F_{{\rm tot},z}$.  The global magnetization agrees
                  very well with the theoretical predictions of
                  Ref.~\cite{Katsura} and behaves in the same way
                  whatever $t/U_0$ and $\rho$, as expected.  }
\label{NEW_magnetic_operator_rho_1_1p5_2_L4b120Utz0p1_tsurU0varie}
\end{center}
\end{figure}

\begin{figure}[!h]
\begin{center}
  \hspace{-0.1 cm}\includegraphics[height=11. cm]{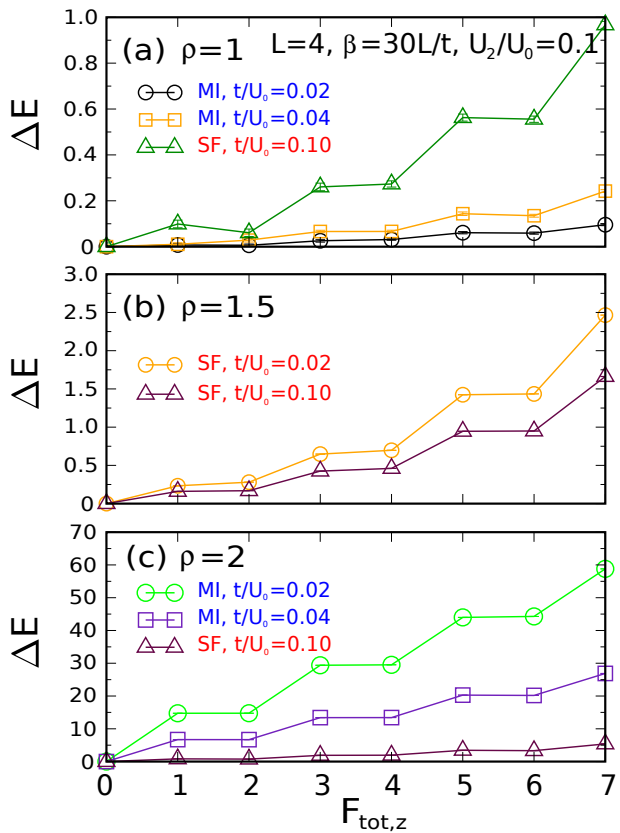}
 \caption{(Color online) The energy difference $\Delta E = E(F_{{\rm
        tot},z})-E(F_{{\rm tot},z}=0)$ versus $F_{{\rm tot},z}$ in the MI and SF
    phases for $\rho=1$ (a), $\rho=1.5$ (b) and $\rho=2$ (c).  The
    energy, minimum at $F_{\rm tot,z}=0$, increases plateau by plateau
    with $F_{{\rm tot},z}$ as predicted in the theorem of
    Ref.~\cite{Katsura}.  }
\label{spinone_AF_E_versus_Fz}
\end{center}
\end{figure}

We now focus on the dependence of the global magnetization and energy on
$F_{{\rm tot},z}=N_+-N_-$.  According to
Ref.~\cite{Katsura}, the ground state is unique, for all values of
the parameters, and has a global magnetization $F_{\rm tot}^2 = F_{\rm
  tot}(F_{\rm tot}+1)$ with quantum number $F_{\rm tot}=|F_{{\rm
    tot},z}|$ if $F_{{\rm tot},z}$ is even and $F_{\rm tot}=|F_{{\rm
    tot},z}|+1$ if $F_{{\rm tot},z}$ is odd (we focus on the even
$N$ case).
Fig. \ref{NEW_magnetic_operator_rho_1_1p5_2_L4b120Utz0p1_tsurU0varie}
shows that the global magnetization $F^2_{\rm tot}$ agrees very well
with the theoretical predictions of Ref.~\cite{Katsura} and $F^2_{\rm
  tot}$ behaves in the same way whatever $t/U_0$ and $\rho$, as
expected.
According to Ref.~\cite{Katsura}, the ground state energy satisfies
$E(F_{{\rm tot},z})<E(F_{{\rm tot},z}+1)$ if $F_{{\rm tot},z}$ is even
and $E(F_{{\rm tot},z})=E(F_{{\rm tot},z}+1)$ if $F_{{\rm tot},z}$ is
odd.  Fig.~\ref{spinone_AF_E_versus_Fz} shows that this result is also
verified in the whole phase diagram.  This confirms that the energy is
minimized when the spins are in the plane $F_{{\rm tot},z}=0$ and that
this constraint used in the rest of the study allows us to reach the
ground state.

 \subsection{Phase transitions}

\begin{figure}[!h]
   \includegraphics[height= 6.6 cm]{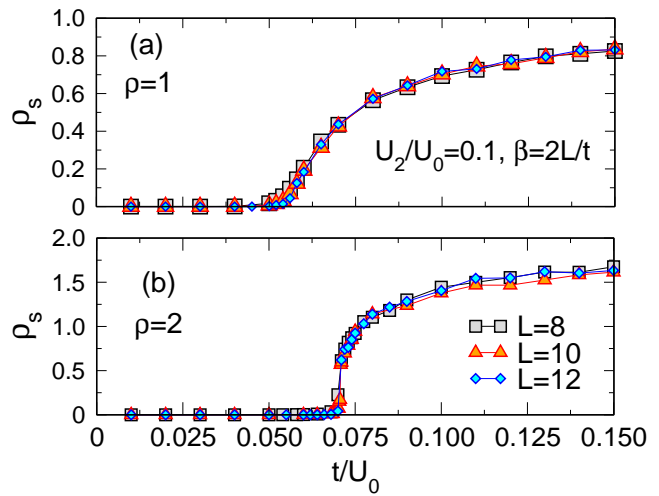}
 \caption {(Color online) The superfluid density as a function of
   $t/U_0$ at fixed total particle density for $U_2/U_0 = 0.1$. 
 The transition from the MI (small $t/U_0$) to the SF phase is
 signaled by $\rho_s$ becoming finite.  For $\rho=1$ (a) the
 transition is continuous whereas the jump in $\rho_s$ marks the
 presence of a first-order phase transition for $\rho=2$ (b). 
}
 \label{spinone_AF_QMC_transition_rho1et2_Utz0p1}
 \end{figure}

 An important difference between one and two dimensions is that
 first-order phase transitions may appear in the latter whereas they
 are rather rare in the former \cite{batrouni2009}.  Since
 MFT predicts a first-order phase transition for even Mott lobes
 \cite{tsuchiya05, pai08}, we used QMC simulations to study this
 transition in detail.
 Figure~\ref{spinone_AF_QMC_transition_rho1et2_Utz0p1} shows the
 evolution of the superfluid density, $\rho_s$, at the MI-SF transition
 at fixed total density while varying $t/U_0$.  The MI-SF transition
 with $\rho=1$ is continuous
 (Fig.~\ref{spinone_AF_QMC_transition_rho1et2_Utz0p1} (a)), whereas it
 is discontinuous with $\rho=2$
 (Fig.~\ref{spinone_AF_QMC_transition_rho1et2_Utz0p1} (b)): the jump
 in the superfluid density indicates a first-order phase transition.
 In Fig.~\ref{spinone_AF_QMC_transition_rho1et2_Utz0p1}, the finite
 size effects are small and the MI-SF transition with
 $\rho=1$ is continuous 
 which disagrees with Ref.~\cite{toga} where a first order
phase transition is observed. The transition should be in the universality
class of the three dimensional XY model \cite{fisher}.

\begin{figure}[!h]
\begin{center}
  \hspace{1.0 cm}\includegraphics[height=6.6 cm]{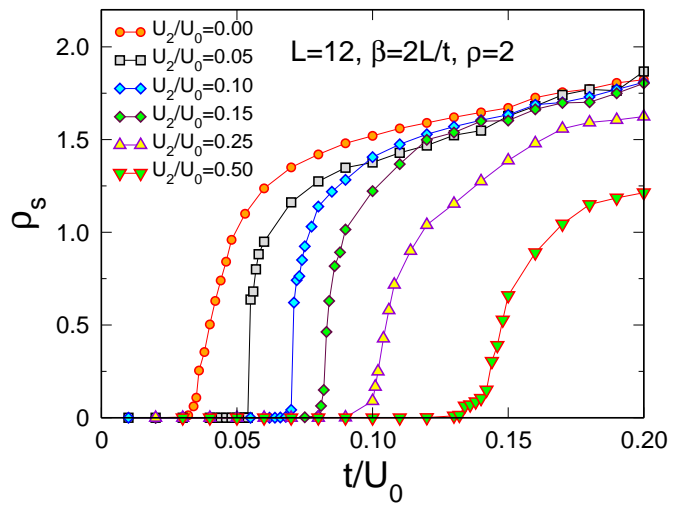}
  \caption{(Color online) The superfluid density as a function of
    $t/U_0$ for $\rho=2$ and for several positive values of $U_2/U_0$.
    The jump in $\rho_s$ indicate the presence of a first-order
    transition at the tip of the $\rho=2$ Mott lobe for small but
    finite values of $U_2/U_0$.  The jumps increase from 0 at
    $U_2/U_0=0$ to a maximum for $U_2/U_0\simeq0.05$ and then decrease
    back to zero for $U_2/U_0\geqslant0.15$.}
\label{spinone_AF_rhos_QMC_transition_rho2_Utzvarie}
\end{center}
\end{figure}

 The jump in $\rho_s$ at the MI-SF phase transition with $\rho=2$ is
 observed for different values of $U_2/U_0$.
 Fig.~\ref{spinone_AF_rhos_QMC_transition_rho2_Utzvarie} shows that
 the transition is continuous for $U_2/U_0=0$, then it becomes
 discontinuous for small values, $U_2/U_0 < 0.15$, and is again
 continuous for $U_2/U_0 \geqslant 0.15$.  The jump in the superfluid
 density varies continuously from zero at $U_2/U_0=0$ to a maximum for
 $U_2/U_0 \simeq 0.05$ and then decreases back to zero as $U_2/U_0$ is
 increased further.  This interval disagrees with the variational study
 \cite{tsuchiya05} which predicted the first-order transition for
 $U_2/U_0$ smaller than 0.32.  A similar behavior was observed for a spin
 1/2 model in Ref. \cite{deforges11}.
\begin{figure}[!h]
\begin{center}
 	\hspace{-0.1 cm}\includegraphics[height=6.6 cm]{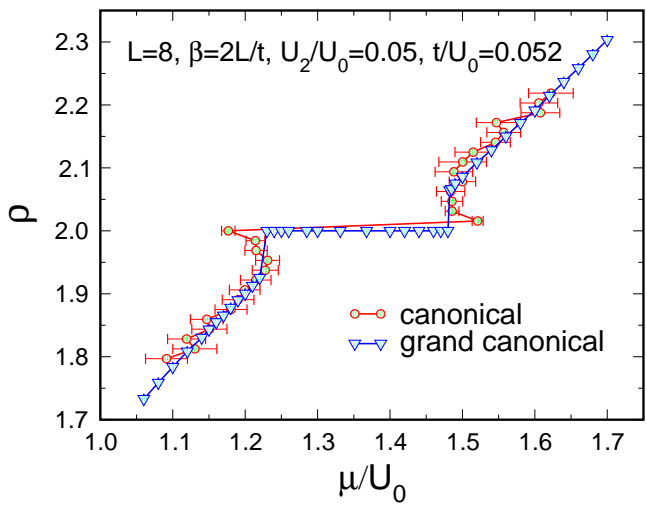} 
        \caption {(Color online) The density, $\rho$, as a function of
          the chemical potential, $\mu$, close to the tip of the
          $\rho=2$ Mott lobe with $U_2/U_0=0.05$.  The canonical
          simulation shows a negative compressibility region, $\kappa
          \propto \partial \rho/\partial \mu<0$.  The grand canonical
          simulation exhibits a corresponding jump in the
          density. Thus both approaches show the presence of a
          first-order transition.}
 \label{spinone_rhovsmu_CetGC_tUz0p05Ut0p052Uz}
 \end{center}
 \end{figure}

 To confirm the presence of first-order phase transitions near the tip
 of the MI lobe for even densities, we studied the behavior of $\rho$
 as a function of $\mu$ as one cuts across the lobe at fixed $t/U_0$,
 using QMC simulations in the canonical and grand canonical ensembles.
 In the canonical ensemble, a first-order transition is signaled by
 negative compressibility \cite{batrouni2000}, $\kappa
 \propto \partial \rho/\partial \mu<0$.  In the grand canonical
 ensemble, there will be a corresponding discontinuous jump in the
 $\rho$ versus $\mu$ curve.
 Figure~\ref{spinone_rhovsmu_CetGC_tUz0p05Ut0p052Uz} shows both these
 cases.  The canonical simulations clearly show negative $\kappa$ just
 before and after the Mott plateau at $\rho=2$.  On the other hand,
 the grand canonical ensemble shows discontinuous jumps in $\rho$ at
 the corresponding values of $\mu$.  The canonical and grand canonical
 simulations are in quantitative agreement on the size of the unstable
 region, which is extremely narrow; the system is stable for densities
 smaller
 than $\rho=1.92$ or larger than $\rho=2.06$ for the chosen value of
 $t/U_0=0.052$ with $U_2/U_0=0.05$.\\ 

\begin{figure}[!h]
\begin{center}
  \hspace{0.0 cm}\includegraphics[height= 6.2 cm]{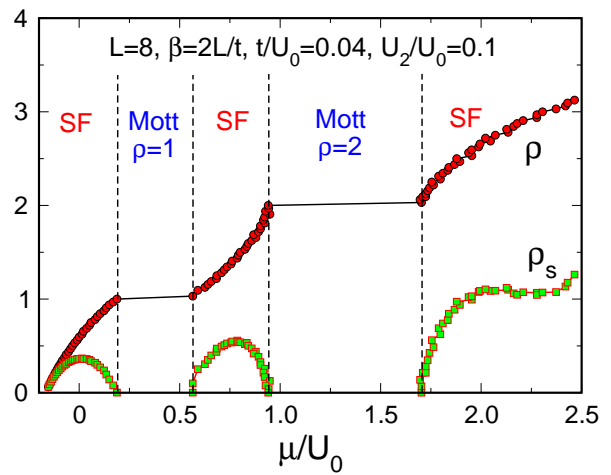}
  \caption{(Color online) The density, $\rho$, versus chemical
    potential, $\mu$, exhibits the usual Mott plateaux at commensurate
    filling for $U_2/U_0=0.1$. Also shown is the superfluid density
    $\rho_s$ in the compressible phases.  Both $\rho$ and $\rho_s$ are
    continuous as the MI phase is approached: the transitions are
    second order.  }
\label{spinone_rhovsmu_MCQ_AF_coupediagphase_Utz0p1_tsurUz_0p040_L8beta16}
\end{center}
\end{figure}

 Figure
\ref{spinone_rhovsmu_MCQ_AF_coupediagphase_Utz0p1_tsurUz_0p040_L8beta16}
 shows a vertical slice of the phase diagram for $t/U_0=0.04$ 
 (Fig.~\ref{Spinone_QMC_diagphase_antiferro_Ut0p1Uz}).  This slice goes
through the first two incompressible Mott plateaux ($\kappa
 \propto \partial \rho /\partial \mu=0$ and $\rho_s=0$) but remains
 outside of the $\rho=3$ Mott phase.  MFT predicts that the MI-SF
 transition near the tip of the even Mott lobes is first order
 \cite{pai08,tsuchiya05}.  For $t/U_0=0.04$, $\rho_s$ vanishes continuously as a
 function of $\mu$ as the MI phase is approached and the
 compressibility is positive for all $\mu$, indicating that the MI-SF
 transitions are second order.  Nevertheless, $|\kappa| \propto
 |\partial \rho /\partial \mu|$, increases to a huge value as the
 $\rho=2$ MI phase is approached.  This behavior is not observed as the
 $\rho=1$ MI phase is approached and indicates one approaches the
 metastable region close to the tip of the even Mott lobes.

\subsection{Magnetic properties}

\begin{figure}[!h]
\begin{center}
  \includegraphics[height=6.7 cm]{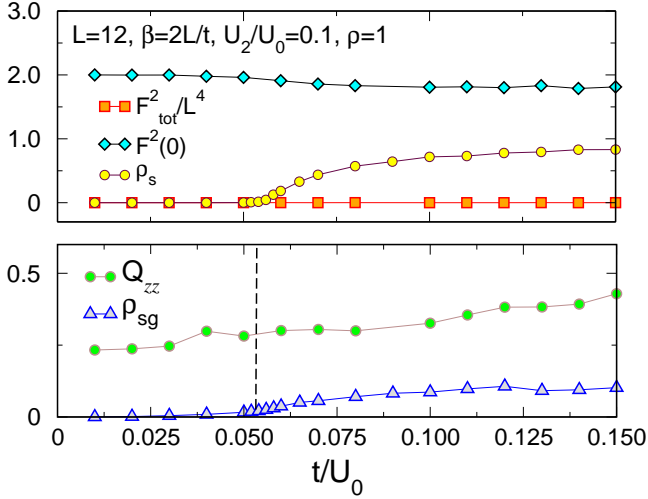}
  \caption{(Color online) Global magnetization $F^2_{\rm tot}$, local
    magnetic moment $F^2(0)$, superfluid density $\rho_s$, nematic
    order parameter $Q_{zz}$ and singlet density $\rho_{\rm sg}$ for
    the MI-SF transition at fixed density $\rho=1$.  The system,
    nematic in the MI phase, remains nematic in the superfluid phase.
The vertical dashed line indicates the position of the MI-SF transition.}
\label{spinone_AF_FF_rhos_QMC_transition_rho1_Utz0p1}
\end{center}
\end{figure}

\begin{figure}[!h]
  \centerline{\includegraphics[width=8.5cm]{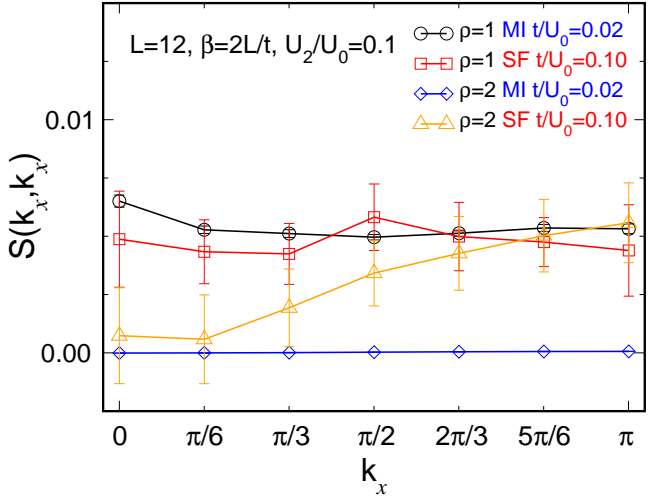}}
  \caption{(Color online) Magnetic structure factor as a function of
    wave vector $k_x$ in different phases. The magnetic structure
    factor always remains negligible. Especially, we do not find a
    peak at ${\bf k} = (\pi,\pi)$ that would signal an
    antiferromagnetic order.\label{structure_factor}}
\end{figure}

We now examine in detail the magnetic properties of the system and
their behavior across the phase transitions.  For $\rho=1$
(Fig.~\ref{spinone_AF_FF_rhos_QMC_transition_rho1_Utz0p1}), the local
moment $F^2(0)=2$ is constant in the Mott lobe, as one have exactly
one particle per site, and slightly decreases to $F^2(0)\simeq1.8$ at
$t/U_0=0.15$ because of the fluctuations of the density in the
superfluid phase.  The global magnetization, equal to zero, is not
affected by the transition.  The nematic order parameter $Q_{zz}$ is
always non zero in the Mott and superfluid phases, which
indicates a spin anisotropy in these phases. At the same time, the magnetic
structure factor $S({\bf k})$
(Fig.~\ref{structure_factor}) shows no trace of magnetic order
anywhere in the phase diagram, especially no sign of antiferromagnetic
spin order that would be compatible with zero total magnetization.
We, therefore, conclude that there is nematic order in these two phases.
Figure~\ref{spinone_AF_FF_rhos_QMC_transition_rho1_Utz0p1} also shows
the evolution of the singlet density $\rho_{\rm sg}$
(Eq.~\ref{spinone_singletdensity}) which is zero in the MI phase and
remains weak in the superfluid.

\begin{figure}[!h]
\begin{center}
  \hspace{0.0 cm}\includegraphics[height=6.4 cm]{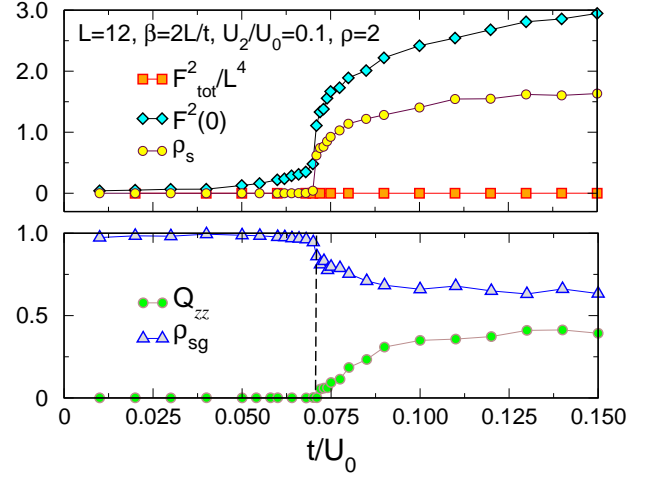}
  \caption{(Color online) Global magnetization $F^2_{\rm tot}$, local
    magnetic moment $F^2(0)$, superfluid density $\rho_s$, nematic
    order parameter $Q_{zz}$ and singlet density $\rho_{\rm sg}$ for
    the MI-SF transition at fixed density $\rho=2$.  The system,
    singlet in the MI phase, becomes nematic in the superfluid phase.
The vertical dashed line indicates the position of the MI-SF transition.
  }
\label{spinone_AF_FF_rhos_QMC_transition_rho2_Utz0p1}
\end{center}
\end{figure}

Whereas the nature of magnetism does not change for the MI-SF
transition with $\rho=1$, things change dramatically for $\rho=2$.  In
Fig.~\ref{spinone_AF_FF_rhos_QMC_transition_rho2_Utz0p1}, the values
of $t/U_0$ take the system from deep in the second Mott lobe
($\rho_s=0$) to the superfluid phase ($\rho_s\neq0$).  As shown for
$\rho=1$ (Fig.~\ref{spinone_AF_FF_rhos_QMC_transition_rho1_Utz0p1}),
the global magnetization $F^2_{\rm tot}$ is zero all along the
transition.  Deep in the MI lobe, the system has only singlets: $F^2(0)=0$
with $\rho_{\rm sg}=1$, while it is nematic in the superfluid phase:
$F^2(0)$ and $Q_{zz}\neq0$. In spite of the nematic order in the
superfluid phase, pairs of singlet are still present: $\rho_{\rm
  sg}\neq0$.  The precision of our results does not allows us to
study the coherence of singlet pairs.  Thus, we cannot
conclude if the superfluidity is carried by individual particles or by the
singlet pairs. In the Mott
lobe, close to the transition, at $t_c/U_0\simeq 0.071$, the local
magnetic moment $F^2(0)$ increases because of the quantum
fluctuations, whereas the spin remains isotropic: $Q_{zz}=0$. At the
transition, simulations show discontinuous jumps in both $F^2(0)$ and
$\rho_s$. The spin anisotropy, $Q_{zz}\neq0$, grows continuously from
zero in the superfluid close to the transition.

\begin{figure}[!h]
\begin{center}
\includegraphics[height=6.4 cm]{fig22a.eps}
\includegraphics[height=6.4 cm]{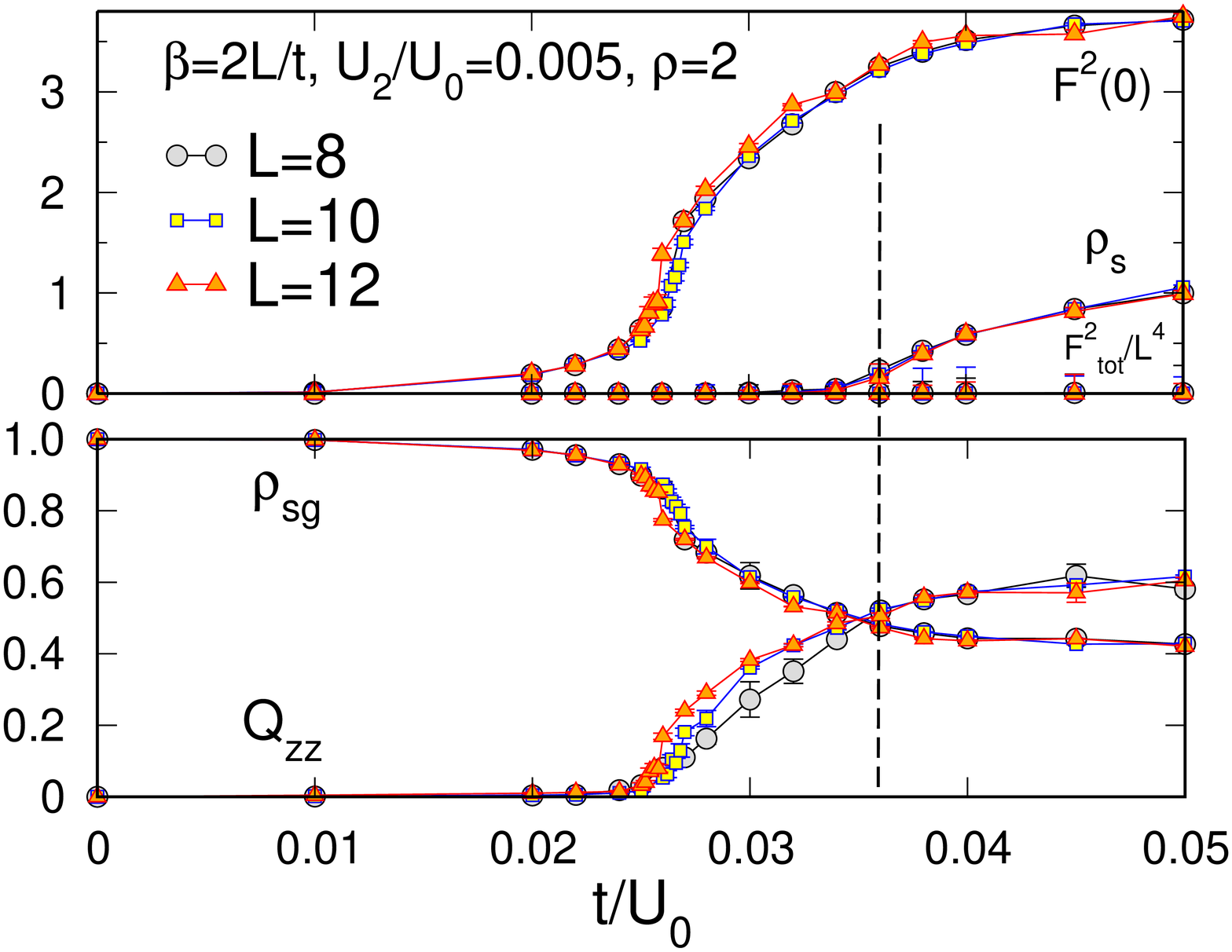}
\caption{(Color online) Global magnetization $F^2_{\rm tot}$, local
  magnetic moment $F^2(0)$, superfluid density $\rho_s$, nematic order
  parameter, $Q_{zz}$, and singlet density $\rho_{\rm sg}$ for the
  MI-SF transition at fixed density $\rho=2$ for several system sizes
  $L$ and for two values of $U_2/U_0$: $U_2/U_0=0.01$ (top) and
  $U_2/U_0=0.005$ (bottom).  Deep in the Mott lobe, the system is
  singlet whereas the system is nematic in the superfluid phase.  The
  singlet-nematic transition seems to appear in the MI lobe, close to
  the transition, as $L$ is increased.  In the case where
  $U_2/U_0=0.005$, the simulation on the largest $L=12$ size shows a
  discontinuity in the evolution of the physical quantities, thus
  indicating a possible first order transition between a singlet and a
  nematic Mott insulator phase. The vertical dashed lines indicate the
positions of the MI-SF transitions. }
\label{spinone_AF_L8_10_12_QMC_transition_rho2_Utz0p01}
\end{center}
\end{figure}

According to Refs.~\cite{imambekov04, demler02, snoek04}, for insulating states
with an even number of particles per site, there is always a spin
singlet phase, but there may also be a first order transition between
a singlet and a nematic phase inside the $\rho=2$ Mott phase for small
values of $U_2/U_0 < 0.025$ in two dimensions.  In
Fig.~\ref{spinone_AF_L8_10_12_QMC_transition_rho2_Utz0p01}, we
elucidate this question by using smaller values of $U_2/U_0$: 0.01 and
0.005.  As observed for $U_2/U_0=0.10$, the system is singlet deep in
the Mott lobe and nematic in the superfluid phase.
Figure~\ref{spinone_AF_L8_10_12_QMC_transition_rho2_Utz0p01} (top) shows
that the SF-MI transition, $\rho_s\to0$, occurs at a larger $t/U_0$
than the transition to nematic order $Q_{zz}\to 0$ for
$U_2/U_0=0.01$.  Thus the singlet-nematic transition seems to take
place inside the MI phase, but quite close to the MI-SF
transition. More interestingly, we see that the position of the
singlet-nematic transition is shifted to the left with increasing
sizes whereas the position of the MI-SF transition is almost
unchanged, which gives a clearer separation of the two transition
points for larger sizes.  For example, for $L=12$, the MI-SF
transition happens for $t/U_0 \simeq 0.037$, whereas the nematic
transition happens for $t/U_0 \simeq 0.031$.  This effect is more
pronounced in the second case $U_2/U_0=0.005$
(Fig.~\ref{spinone_AF_L8_10_12_QMC_transition_rho2_Utz0p01} (bottom))
where the transitions take place at $t/U_0 \simeq 0.037$ and $t/U_0
\simeq 0.025$ respectively and where the nematic region in the Mott
phase is much wider.  
In both these cases, it is difficult to conclude
if there is a first or a second order MI-SF transition. We observe
a seemingly continuous transition but we are close to 
$U_2/U_0=0$ where the transition is continuous. If there is
a discontinuity, we would except the jump in $\rho_s$ to be very small and, hence, 
very difficult to observe.
In the $U_2/U_0=0.01$ case, there is
also no evidence
of a first-order transition for the singlet-nematic transition, as
$F^2(0)$ and $Q_{zz}$ evolve in a seemingly continuous way with our
precision (Fig.~\ref{spinone_AF_L8_10_12_QMC_transition_rho2_Utz0p01}
(top)). However, a discontinuity is observed in these quantities
(Fig.~\ref{spinone_AF_L8_10_12_QMC_transition_rho2_Utz0p01} (bottom))
for the largest size, $L=12$, for $U_2/U_0=0.005$; this signals a
discontinuous transition. A rigorous investigation
with larger sizes $L>12$, currently inaccessible with QMC, would be
required to specify the exact point where the transition occurs and to
see if this discontinuous transition is also present in the $U_2/U_0=0.01$ case.
We have also studied the $U_2/U_0=0.05$ case but it was impossible to
distinguish the MI-SF phase transition from the singlet-nematic one, which
then appears similar to the $U_2/U_0 = 0.10$ case.
 
\begin{figure}[!h]
\begin{center}
  \includegraphics[width=8.5cm]{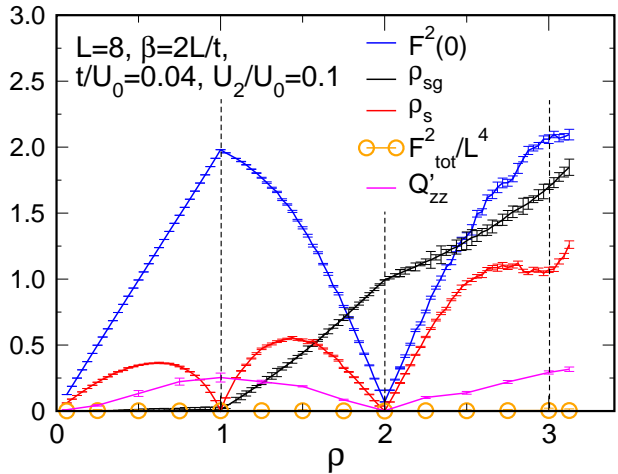}
  \caption {(Color online) Local magnetic moment $F^2(0)$, singlet
    density, $\rho_{\rm sg}$, superfluid density, $\rho_s$, global
    magnetization $F^2_{\rm tot}$, and nematic order parameter
    $Q'_{zz}$ versus the density $\rho$ associated with the vertical
    slice at $t/U_0=0.04$ of the phase diagram
    Fig.~\ref{Spinone_QMC_diagphase_antiferro_Ut0p1Uz}. }
\label{spinone_rhos_MCQ_AF_coupediagphase_Utz0p1_tsurUz_0p040_L8beta16}
\end{center}
\end{figure}

Figure~\ref{spinone_rhos_MCQ_AF_coupediagphase_Utz0p1_tsurUz_0p040_L8beta16}
shows the local magnetic moment, $F^2(0)$, the singlet density,
$\rho_{\rm sg}$, the superfluid density, $\rho_s$, the nematic order
parameter, $Q'_{zz}$, and the global magnetization, $F^2_{\rm tot}$,
versus the density taken from
Fig.
\ref{spinone_rhovsmu_MCQ_AF_coupediagphase_Utz0p1_tsurUz_0p040_L8beta16}.
All along this slice, the global magnetization $F^2_{\rm tot}$ is
zero, as expected.  As the density is increased, we see that the
number of singlet increases regularly for $\rho \ge 1$ as
expected. The behavior of the local magnetic moment can be understood
in terms of a minimization process. For an even number of particles on
a site, the minimum value of the local moment is zero whereas it is
$F^2(0)=2$ for an odd number of particles. We then observe
this alternation of increases of the local moment when the number of
sites with an odd number of particles is increasing ($\rho < 1$ or $2
< \rho < 3$) and of decreases when the number of sites with an even
number is increasing ($ 1 < \rho < 2$).  In this case, we measured
$Q'_{zz}$ because $Q_{zz}$ is ill behaved for $\rho<1$, as explained
in section \ref{section2}. We find that the spin anisotropy and nematic
order is present in the whole superfluid phase, whatever the density.
It is also present, as seen before, in odd density Mott phases and
essentially absent from even density Mott phases.


\section{Conclusion}

We studied spin-1 bosons with interactions on a lattice. The
interactions are among the simplest possible for such a system: an
on-site repulsion independent of spin and an on-site ferromagnetic or
antiferromagnetic coupling between spins on the same site.

We have shown that, for both ferromagnetic ($U_2<0$) and
antiferromagnetic ($U_2>0$) interactions, our QMC results are in good
agreement with the theorem of Ref.~\cite{Katsura}.

For $U_2<0$ we showed that all MI-SF transitions are continuous and
that the whole phase diagram is ferromagnetic. This is very similar to
what was observed in the one dimensional case \cite{batrouni2009},
although the nature of the MI-SF transitions with fixed integer
density differs from the 1D case.  The ferromagnetism of the whole
phase diagram was demonstrated by Katsura and Tasaki and we verified
numerically their theorem. We also determined the extent of the
different Mott and superfluid phases and showed that the
magnetic behavior is indeed not affected by the different natures
(superfluid or solid) of theses phases. This is rather surprising as
our model does not include explicit coupling between spins on
different sites. These couplings arise due to the movement of the
particles, and these movements are quite different in solid and
superfluid phases.

For $U_2>0$, the situation is more interesting as Katsura and
Tasaki's theorem only predicts the global value of the magnetization
$F_{\rm tot}^2$ to be zero in the simple case where $F_{{\rm
    tot},z}=0$.  This value is compatible with several magnetic
orders, namely an antiferromagnetic order or a nematic order, and the
theorem does not specify which order, if any, will emerge in the ground
state.  Analysing the magnetic structure factor, we
excluded the possibility of antiferromagnetic order.  The first Mott
lobe was expected to be nematic \cite{imambekov04,toga} via a
mapping on a spin-1 Heisenberg model and, indeed, we found that the
nematic order parameter $Q_{zz}$ was non zero in this phase.  More
surprinsingly, nematic order is also present in the superfluid
phase.  Furthermore, the density distribution differs from that expected
from MFT \cite{pai08} which predicts another possible ground state where
the spin-0 population dominates.  In the second Mott lobe, we observe the
formation of singlet order for $U_2/U_0=0.1$, and a
singlet-nematic transition inside the Mott lobe, close to the tip
of the lobe with $U_2/U_0=0.01$ and $U_2/U_0=0.005$ as expected
\cite{imambekov04}. We observed such a transition on the sizes we
could study, and were able to find evidence of a first order
transition for the $U_2/U_0=0.005$ case (although we did not observe
such evidence for $U_2/U_0=0.01$) as predicted with MFT in
\cite{imambekov04, demler02, snoek04}.  Finally, we demonstrate that the MI-SF
transition in the $\rho=1$ lobe is continuous, independently of $U_2$,
whereas the transition in the $\rho=2$ lobe is discontinuous
(first-order) for $0.05 < U_2/U_0 < 0.15$ and continuous for
larger values.

\acknowledgements 
  This work was supported by: the CNRS-UC Davis EPOCAL LIA joint
  research grant; by NSF grant OISE-0952300; and ARO Award
  W911NF0710576 with funds from the DARPA OLE Program.


\begin{thebibliography}{}

\bibitem{bloch08} I. Bloch, J. Dalibard, and W. Zwerger,
  Rev. Mod. Phys. {\bf 80}, 885 (2008).

\bibitem{Greiner02} M. Greiner, O. Mandel, T. Esslinger,
  T.W. H\"ansch, and I. Bloch, Nature {\bf 415}, 39 (2002).

\bibitem{Schneider08} U. Schneider, L. Hackerm\"uller, S. Will,
  Th. Best, I. Bloch, T. A. Costi, R. W. Helmes, D. Rasch, and
  A. Rosch, Science {\bf 322}, 5907 (2008).

\bibitem{Jordens08} R. J\"ordens, N. Strohmaier, K. G\"unter,
  H. Moritz, and T. Esslinger, Nature {\bf 455}, 204 (2008).

\bibitem{Stamper1998} D. M. Stamper-Kurn, M. R. Andrews,
  A. P. Chikkatur, S. Inouye, H.-J. Miesner, J. Stenger, and
  W. Ketterle, Phys. Rev. Lett. {\bf 80}, 2027 (1998).

\bibitem{Vengalattore08} M. Vengalattore, S. R. Leslie, J. Guzman, and
  D. M. Stamper-Kurn, Phys. Rev. Lett. {\bf 100}, 170403 (2008).

\bibitem{Vengalattore10} M. Vengalattore, J. Guzman, S. R. Leslie,
  F. Serwane, and D. M. Stamper-Kurn, Phys. Rev. A {\bf 81}, 053612
  (2010).

\bibitem{ho} T.-L. Ho, Phys. Rev. Lett. {\bf 81}, 742 (1998).

\bibitem{ohmi98} T. Ohmi and K. Machida, J. Phys. Soc. Jpn. {\bf 67}, 1822 (1998).


\bibitem{Theis} M. Theis, G. Thalhammer, K. Winkler, M. Hellwig,
  G. Ruff, R. Grimm, and J. H. Denschlag, Phys. Rev. Lett. {\bf 93},
  123001 (2004); G. Thalhammer, M. Theis, K. Winkler, R. Grimm, and
  J. H. Denschlag, Phys. Rev. A {\bf 71}, 033403 (2005).



\bibitem{Stamperbook} D. M. Stamper-Kurn and W. Ketterle, in
  \textit{Coherent Atomic Matter Waves}, edited by R. Kaiser,
  C. Westbrook, and F. David (Springer, Berlin, 2001), p. 137.

\bibitem{imambekov04} A. Imambekov, M. Lukin, and E. Demler,
  Phys. Rev. A {\bf 68}, 063602 (2003) and Phys. Rev. Lett. {\bf 93},
  120405 (2004).


\bibitem{Kawashima02} N. Kawashima, \textit{Prog. Theor. Phys. Suppl.}
  {\bf 145}, 138 (2002).

\bibitem{Tsuchiya04} S. Tsuchiya, S. Kurihara, and T. Kimura,
  Phys. Rev. A {\bf 70}, 043628 (2004).


\bibitem{pai08} R.V. Pai, K. Sheshadri, and R. Pandit, Phys. Rev. B {\bf
    77}, 014503 (2008).

\bibitem{tsuchiya05} T. Kimura, S. Tsuchiya, and S. Kurihara,
  Phys. Rev. Lett. {\bf 94}, 110403 (2005).

\bibitem{toga} Y. Toga, H. Tsuchiura, M. Yamashita, K. Inaba, and
  H. Yokoyama, J. Phys. Soc. Jpn. {\bf 81}, 063001 (2012).

\bibitem{Katsura} H. Katsura and H. Tasaki, Phys. Rev. Lett. {\bf
    110}, 130405 (2013).


\bibitem{Kimura13}
T. Kimura, Phys. Rev. A {\bf 87}, 043624 (2013).

\bibitem{Rizzi05_dmrg} M. Rizzi, D. Rossini, G. De Chiara,
  S. Montangero, and R. Fazio, Phys. Rev. Lett. {\bf 95}, 240404
  (2005).

\bibitem{Bergkvist06_dmrg} S. Bergkvist, I.P. McCulloch, and
  A. Rosengren, Phys. Rev. A {\bf 74}, 053419 (2006).



\bibitem{apaja06} V. Apaja and O. F. Sylju{\aa}sen, Phys. Rev. A {\bf
    74}, 035601 (2006).

\bibitem{batrouni2009} G. G. Batrouni, V. G. Rousseau, and
  R. T. Scalettar, Phys. Rev. Lett. {\bf 102}, 140402 (2009).

\bibitem{demler02} E. Demler and F. Zhou, Phys. Rev. Lett. {\bf 88},
  163001 (2002).
\bibitem{snoek04} M. Snoek and F. Zhou, Phys. Rev. B {\bf 69}, 094410
  (2004).

\bibitem{deforges10} L. de Forges de Parny, M. Traynard, F. H\'ebert,
  V. G.  Rousseau, R. T. Scalettar, and G. G. Batrouni, Phys. Rev. A
  {\bf 82}, 063602 (2010).

\bibitem{deforges11} L. de Forges de Parny, F. H\'ebert,
  V. G. Rousseau, R .T.  Scalettar, and G. G. Batrouni, Phys. Rev. B
  {\bf 84}, 064529 (2011).

\bibitem{kruti04} 
K. V. Krutitsky and R.Graham, Phys. Rev. A {\bf 70}, 063610 (2004). 

\bibitem{kruti05} 
K. V. Krutitsky, M.Timmer, and R.Graham, Phys. Rev. A {\bf 71}, 033623 (2005).

\bibitem{SGF} V. G. Rousseau, Phys. Rev. E {\bf 77}, 056705 (2008).

\bibitem{directedSGF} V. G. Rousseau, Phys. Rev. E {\bf 78}, 056707
  (2008).

\bibitem{roy} D. M. Ceperley and E. L. Pollock, Phys. Rev. B {\bf 39},
  2084 (1989).

\bibitem{Kuklov2003} A. B. Kuklov and B. V. Svistunov,
  Phys. Rev. Lett. {\bf 90}, 100401 (2003).

\bibitem{svistunov2} B. Capogrosso-Sansone, S. G. S\"oyler,
  N. V.~Prokof\'ev and B. V. Svistunov, Phys. Rev. A {\bf 81}, 053622
  (2010).

\bibitem{fisher} M. P. A. Fisher, P. B. Weichman, G. Grinstein, and
  D. S. Fisher, Phys. Rev. B {\bf 40}, 546 (1989).

\bibitem{batrouni2000} G. G. Batrouni and R. T. Scalettar,
  Phys. Rev. Lett. {\bf 84}, 1599 (2000).



\end{thebibliography}
\end{document}